\documentclass[namedreferences]{solarphysics}
\usepackage[optionalrh,solaenum]{spr-sola-addons}	
\usepackage{graphicx}		
\usepackage{amssymb}	
\usepackage{color}		
\usepackage{url}		
\usepackage{rotating}
 
\usepackage{amsmath}	

\newcommand{\etal}{{\it et al.}}

\newcommand{\eventname}{{\tt SOL2002-09-10T14:53}}
\newcommand{\deltalf}{\delta_{l\!f}}
\newcommand{\deltahf}{\delta_{h\!f}}
\newcommand{\alfahf}{\alpha_{h\!f}}

\newcommand{\aap}{ {\it Astron. Astrophys.}}
\newcommand{\aaps}{ {\it Astron. Astrophys. Suppl.}}

\newcommand{\apj}{ {\it Astrophys. J.}}

\newcommand{\apjl}{ {\it Astrophys. J.}}

\newcommand{\solphys}{{\it Solar Phys.}}

\newcommand{\ssr}{ {\it Space Sci. Rev.}}

\newcommand{\bs}{\begin{sideways}}
\newcommand{\es}{\end{sideways}}
\newcommand\arcmin{\mbox{$^{\prime}$}}
\newcommand\arcsec{\mbox{$^{\prime\prime}$}}

\begin{document}

\begin{article}

\begin{opening}

\title{A burst with double radio spectrum observed up to 212~GHz}

\author{C.G.~\surname{Gim{\'e}nez de Castro}$^{1,2}$\sep
	G.D.~\surname{Cristiani}$^{2,3}$\sep
	P.J.A.~\surname{Sim\~oes}$^{6}$\sep
	C.H.~\surname{Mandrini}$^{2,3}$\sep
        E.~\surname{Correia}$^{1,4}$\sep
	P.~\surname{Kaufmann}$^{1,5}$
}
\runningauthor{C.G. Gimenez de Castro \etal}

\institute{
	$^{1}$ CRAAM, Universidade Presbiteriana Mackenzie, 01302-907, S\~ao Paulo, Brasil;
		email:~\url{guigue@craam.mackenzie.br};
		~\url{paulo@craam.mackenzie.br};
		~\url{kaufmann@craam.mackenzie.br}\\
	$^{2}$ Instituto de Astronom\'{\i}a y F\'{\i}sica del Espacio, 
                CONICET-UBA, CC. 67 Suc. 28, 1428, Buenos Aires, Argentina;
		email:~\url{gcristiani@iafe.uba.ar};
		~\url{mandrini@iafe.uba.ar}\\
	$^{3}$ Facultad de Ciencias Exactas y Naturales, FCEN-UBA, Buenos Aires, Argentina\\
        $^{4}$  Instituto Nacional de Pesquisas Espaciais, S\~ao Jos\'e dos Campos, Brazil\\
	$^{5}$ Centro de Componentes Semicondutores, Universidade Estadual de Campinas, 
	Campinas, Brasil\\
        $^{6}$  School of Physics \& Astronomy, University of Glasgow, Glasgow, Scotland\\
}

\begin{abstract}
We study a solar flare that occurred on September 10, 2002, in active region 
NOAA 10105 starting around 14:52 UT and lasting approximately 5 minutes 
in the radio range. The event was classified as M2.9 in X-rays and 1N in H$\alpha$. 
Solar Submillimeter Telescope observations, in addition to microwave data 
give us a good spectral coverage between 1.415 and 212 GHz. We combine these data 
with ultraviolet images, hard and soft X-rays observations and full-disk magnetograms. 
Images obtained from Ramaty High Energy Solar Spectroscopic Imaging  
data are used to identify the locations of X-ray sources 
at different energies and to determine the X-ray spectrum, while ultra violet images allow us to 
characterize the coronal flaring region. The magnetic field evolution of the active region is analyzed 
using Michelson Doppler Imager magnetograms. The burst is detected at all available radio-frequencies.
X-ray images (between 12~keV and 300~keV) reveal two compact sources and 212
GHz data, used to estimate the radio source position, show a single compact
source displaced by 25\arcsec\ from one of the hard X-ray footpoints. We model the
radio spectra using two homogeneous sources, and combine this analysis with
that of hard X-rays to understand the dynamics of the particles.  Relativistic particles,
observed at radio wavelengths above 50~GHz, have an electron index
evolving with the typical {\em soft--hard--soft} behaviour.  
\end{abstract}

\keywords{Radio Bursts, Association with Flares; Radio Bursts, Microwave; 
X-Ray Bursts, Association with Flares; Flares, 
Relation to Magnetic Field; Chromosphere, Active}

\end{opening}

\section{Introduction}

High frequency radio observations, above 50~GHz,  bring information about
relativistic particles (see e.g. \opencite{Ramatyetal:1994} and 
\opencite{Trottetetal:1998}). Moreover, the efficiency of synchrotron emission,
responsible for the radio radiation, increases as the electron energy
increases, contrary to the bremsstrahlung mechanism which is the origin 
of the Hard X-ray (HXR) emission \cite{Whiteetal:2011}. This makes observations
at high frequencies very attractive for the analysis of high energy
particles.  For typical magnetic fields on the Chromosphere and mildly 
relativistic electrons, gyrosynchrotron theory expects a peak frequency 
at approximately 10~GHz. Therefore the caveat of submillimeter observations 
is that flare emission becomes weaker as the observing frequency increases.  
At the same time, at high frequencies, earth atmosphere becomes brighter 
and absorbs much of the incoming radiation. Notwithstanding some X-class flares 
have shown a second spectrum besides the microwaves spectrum,  with an optically 
thick emission at submillimeter frequencies, sometimes described as an {\em upturn} 
(see e.g. \opencite{Kaufmannetal:2004}, \opencite{Silvaetal:2007}, \opencite{Luthietal:2004b}). 
Nonetheless, \inlinecite{Cristianietal:2008} 
found, in a medium size flare, a second radio component peaking around 200~GHz.
We call these cases double radio spectrum events.\\

Although different mechanisms were proposed to explain the double radio
spectrum events \cite{KaufmannRaulin:2006,FleishmanKontar:2010}, the
conservative approach of two distinct synchrotron sources can fit 
reasonably well to the observations \cite{Silvaetal:2007,Trottetetal:2008}. 
We note, however, that observations at higher frequencies are needed to completely
determine the radiation mechanism of those events that only show the
optically thick emission of the second component, like in \inlinecite{Kaufmannetal:2004} 
and, because of their strong fluxes, have much stringent requirements.\\

Therefore, the double radio spectrum bursts may represent a kind of events
whose low frequency component is the classical gyrosynchrotron from mildly
relativistic particles peaking around 10~GHz, and the high frequency
component is also synchrotron emission with peak frequency around or 
above 50~GHz, depending on the flare characteristics (in some cases above
400~GHz).\\
 
In this work we present a detailed analysis of a double radio spectrum burst 
occurred during a GOES M class event on September 10, 2002 and observed in 
radio from 1.415 to 212~GHz.  We'll show that the low frequency component is 
well correlated with the HXR observed with RHESSI up to 300~keV, 
hence we can study the dynamics of the mildly relativistic electrons inside 
the coronal loop. On the other hand, the high frequency component, which is 
also well represented by an electron synchrotron source, should be produced 
by a different particle population and, likely, in a different place. \\

We first present the data analyzed, explain the reduction methods and
give the clues that justify the interpretations in Section \ref{sec:observations}. 
The spectral analysis, both at radio and Hard X-rays is the kernel of our 
work, thus it deserves the entire Section \ref{sec:analysis}. We divide the
interpretation of the event in two wide energy bands for the mildly
relativistic and the relativistic particles in Section
\ref{sec:discusion}. The consequences of our interpretations are presented
as our final remarks in Section \ref{sec:fin}.  

\section{Observations}
\label{sec:observations}
\subsection{The data}
\label{sec:data}

The impulsive phase of the solar burst \eventname\ started at 14:52:30 UT, 
in active region (AR) NOAA 10105 (S10E43) and lasted a few minutes. The flare is classified as 
GOES M2.9 in soft X-Rays and 1N in H$\alpha$.  The burst was observed by 
various radiotelescopes around the world: (1) the United States Air Force (USAF) Radio Solar 
Telescope Network (RSTN, \opencite{Guidiceetal:1981}) at 1.415, 2.695, 4.995, 
8.8 and 15.4~GHz with 1~s time resolution; (2) the solar polarimeter at 7~GHz 
of the Itapetinga Observatory with 12~ms time resolution \cite{Kaufmann:1971,Correiaetal:1999}; 
(3) the solar patrol radiotelescopes of the Bern University at 11.8, 19.6, 
35 and 50~GHz with 100~ms time resolution (unfortunately, at that time, 
the channel at 8.4 GHz was not working); (4) the null interferometer at 89.4~GHz 
of the University of Bern with 15~ms time resolution \cite{Luthietal:2004a} and (5) 
the Solar Submillimeter Telescope (SST) at 212 and 405~GHz and 40~ms time resolution \cite{Kaufmannetal:2008}.
All telescopes, except the SST, have beams greater than the solar angular size. The
null interferometer can be used to remove the Quiet Sun contribution. The SST beam
sizes are 4\arcmin\ and 2\arcmin\ at 212 and 405~GHz respectively, and form a focal
array that may locate the centroid position of the emitting source and
correct the flux density for mispointing. Below we comment more on this.
Since there are no overlapping frequencies between the different instruments, 
we have to rely on the own telescope calibration procedures, which were 
successfully verified along their long operation history.  All of them, 
except the SST, claim to determine the flux density with an accuracy of the 
order of 10\%. When applying the multi-beam method \cite{GimenezdeCastroetal:1999} 
to SST data, the accuracy is of the order of 20\%.  We have used these figures 
in the present work.\\

HXR data for this event were obtained with the Reuven Ramaty High Energy
Solar Spectroscopic Imager (RHESSI, \opencite{Linetal:2002}). RHESSI provides
imaging and spectral observations with high spatial (0.5\arcsec) and spectral
(1~keV) resolution in the 3 keV--17~MeV energy range.  
We also used complementary data: extreme--ultraviolet (EUV) images 
from the Extreme ultraviolet Imaging Telescope
(SOHO/EIT, \opencite{Delaboudiniereetal:1995}), and full disk magnetograms
from the Michelson Doppler Imager (SOHO/MDI, \opencite{Scherreretal:1995}).
During September 10, 2002, EIT was working in CME watch mode. Observations
were taken at half spatial resolution and with a temporal cadence of 12 minutes,
which is too low compared to the flare duration to detect any kind of flare
temporal evolution. Early during the day, and up to $\sim$14:48~UT, data
were taken at 195~\AA. At around the flare time, from 15:00~UT, and later,
images were obtained at 304~\AA. There is only one image at this wavelength
displaying the flare brightening.  The Transition Region and Coronal Explorer 
(TRACE) was not observing this AR.\\

\subsection{The photospheric evolution around the flare time}
\label{sec:photosphere}

The active region, where the flare occurred, arrived at the eastern solar
limb on September 7, 2002, as an already mature region. It appeared formed
by an unusually strong leading negative polarity and a weaker dispersed
positive following polarity. AR 10105 is the recurrence of AR 10069
seen on the disk during the previous solar rotation.\\

\begin{figure}[h]
\vbox{
\hbox{
\includegraphics[width=0.47\textwidth]{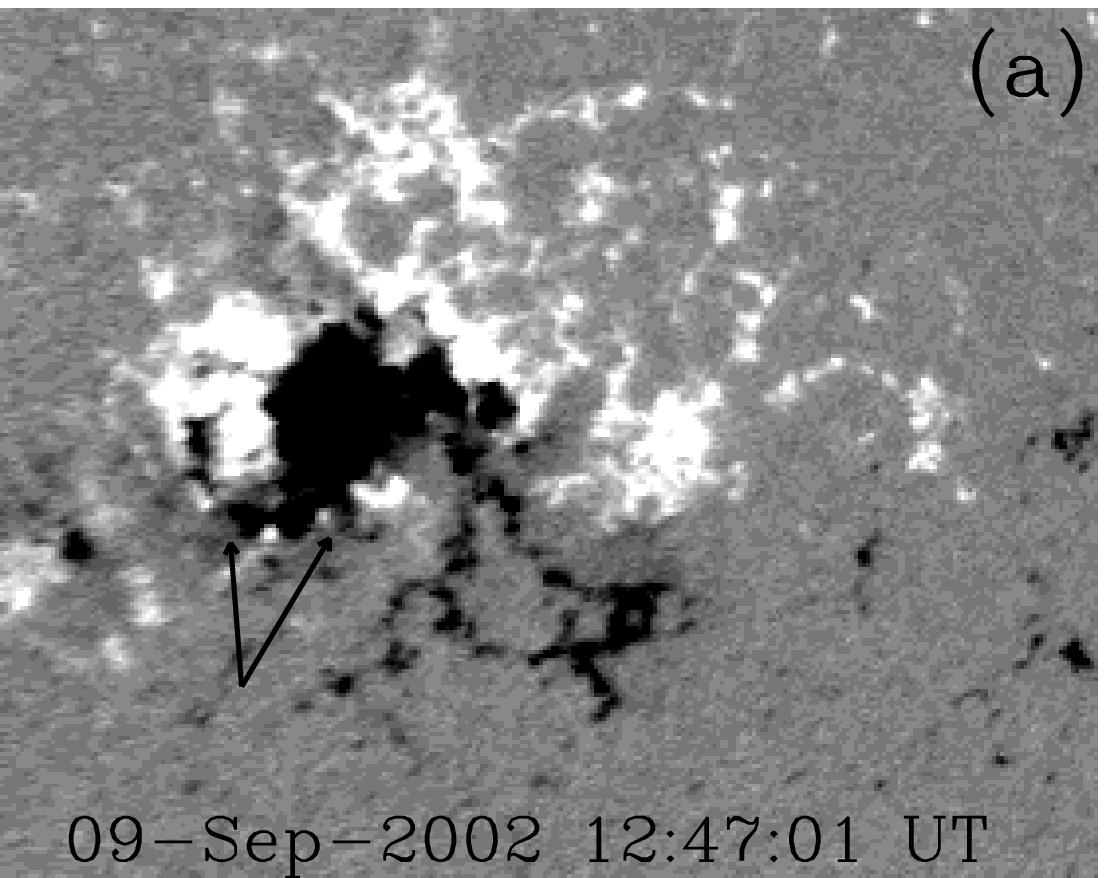}
\hspace{0.12cm}
\includegraphics[width=0.47\textwidth]{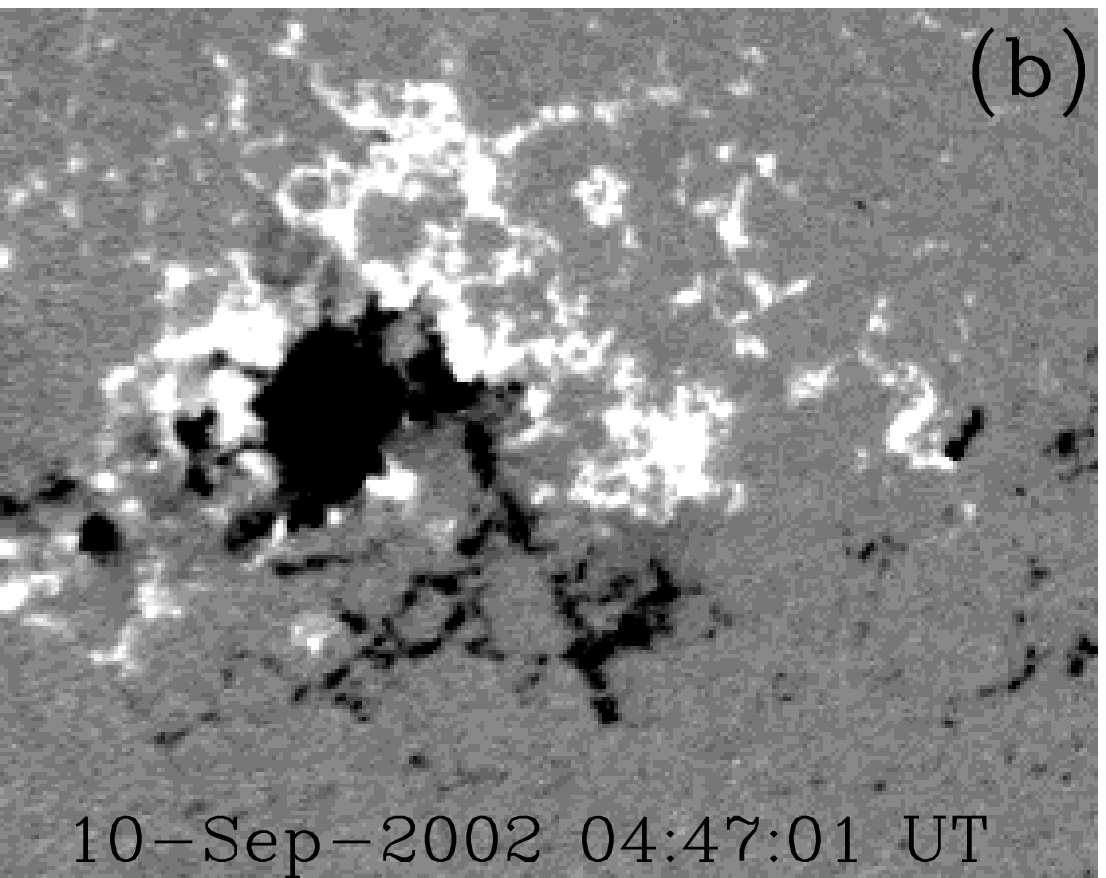}
}
\vspace{0.2cm}
\hbox{
\includegraphics[width=0.47\textwidth]{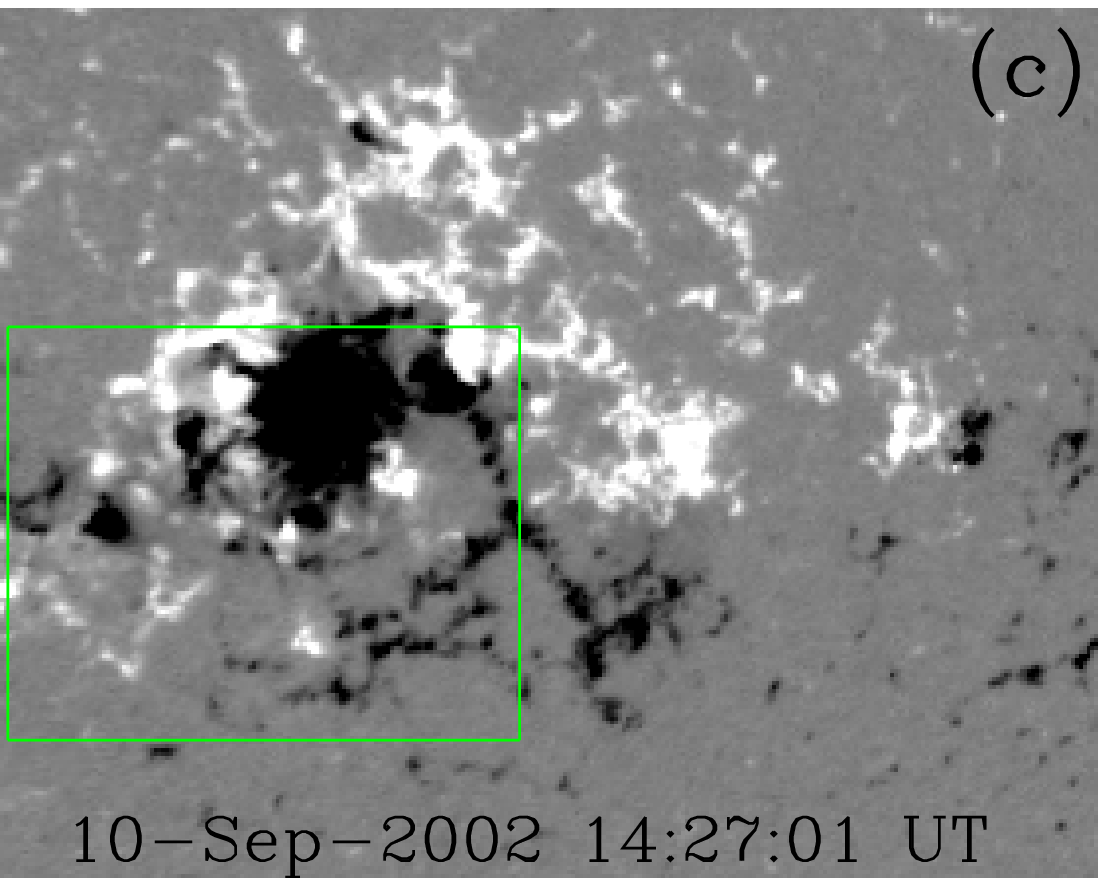}
\hspace{0.12cm}
\includegraphics[width=0.47\textwidth]{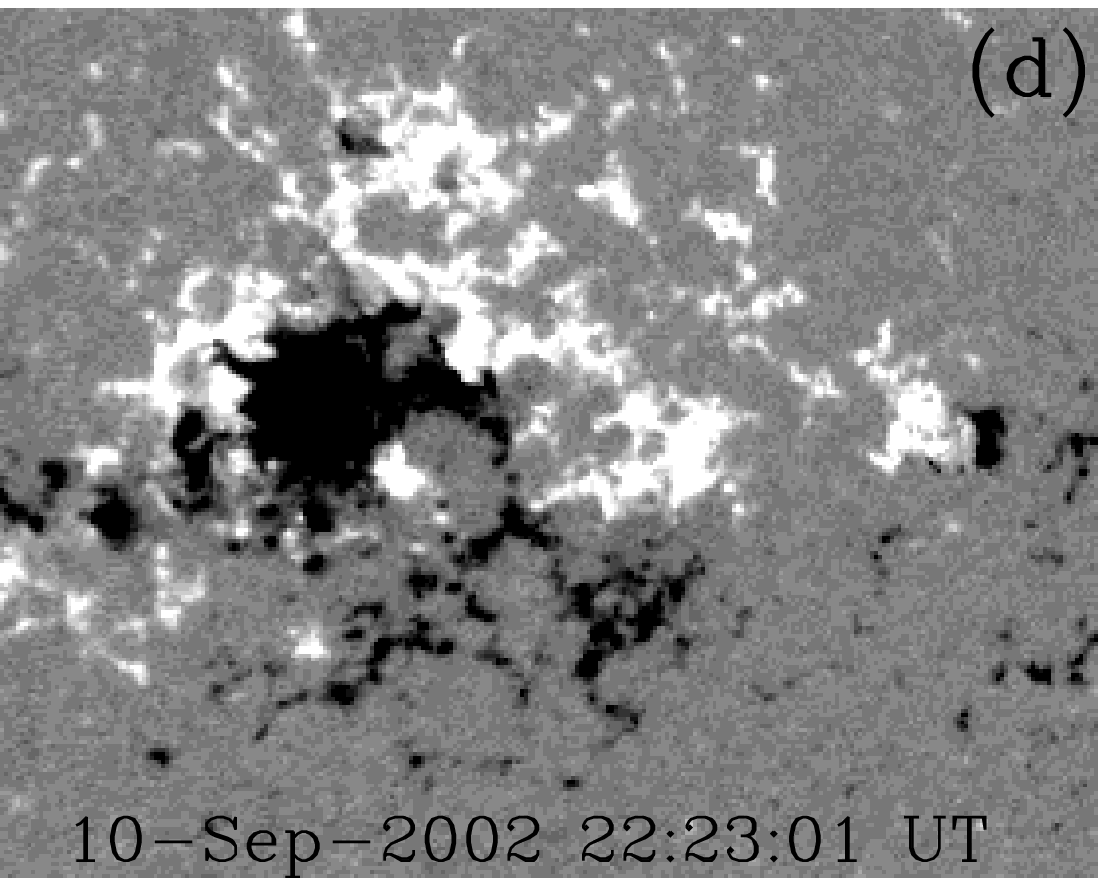}
}
}
\caption{Magnetic field evolution seen in MDI line of sight magnetograms
during September 9 and 10,  2002. All images have been rotated to solar 
Sun center and saturated above (below) 300 G (-300 G). Notice that the 
field intensity in the negative spot is such that saturation is very 
clear in MDI data in Figure~\ref{fig:MDI-HXR-sst}.  The arrows in 
panel (a) point to a chain of small bipoles located in the moat region 
surrounding the spot. The green box in panel (c) represents the 
field of view of the Figure \ref{fig:MDI-HXR-sst}.}
\label{fig:field-evolution}
\end{figure}

Adjacent to AR 10105, on its west, there is a dispersed bipolar
facular region and a new AR to its SE (AR 10108, 
see Figure~\ref{fig:field-evolution}a). Vigorous moving
magnetic feature (MMF, see \opencite{HarveyHarvey:1973})
activity is seen  (compare Figure~\ref{fig:field-evolution}a to
Figure~\ref{fig:field-evolution}b and see the movie 
{\tt mag-field-evol.mpg} that accompanies this paper) around 
the big leading spot and a moat boundary is visible
in Figure~\ref{fig:field-evolution}a mainly in a E-SE segment, 
where magnetic field aggregates at what appears the confluence 
of three supergranular cells. In particular, we have indicated 
within arrows in Figure~\ref{fig:field-evolution}a a chain 
of small bipoles.  Though this image is affected by projection 
effects, these bipolar regions persist in MDI magnetograms.  
Figure~\ref{fig:field-evolution}b shows that the eastern negative 
polarity of the eastermost bipole rotates counter-clockwise 
around the positive one.  This motion should increase the 
magnetic field shear in the region and, at the same time, 
favour the interaction with nearby bipoles in the MMFs to its north.
Figure~\ref{fig:field-evolution}c (see a zoom of this image in
Figure~\ref{fig:MDI-HXR-sst}) shows the magnetic field at 14:27 UT on
September 10, this is the closest in time magnetogram
to the flare occurrence at 14:52 UT. In this magnetogram the negative
polarity to the east (which is along the moat boundary)
has decreased in size as flux cancellation proceeds with the positive
bipole polarity. Along this period, flux cancellation also 
proceeds between this positive polarity and the nearby negative 
polarity of the bipole to the west.  Finally, this small bipole 
is hardly observable by the end of the day (see 
Figure~\ref{fig:field-evolution}d).  The magnetic field evolution 
just discussed lets us infer the origin of the flare eruption and 
the location of the pre-reconnected set of loops and the reconnected 
loops of which only one is visible in the XR and EUV data (see 
Figure~\ref{fig:MDI-HXR-sst} and Section~\ref{fig:MDI-HXR-sst}).\\

\subsection{Time profiles in radio and X-rays}
\label{sec:profiles}

\begin{figure}
\centerline{\resizebox{0.8\textwidth}{!}{\includegraphics{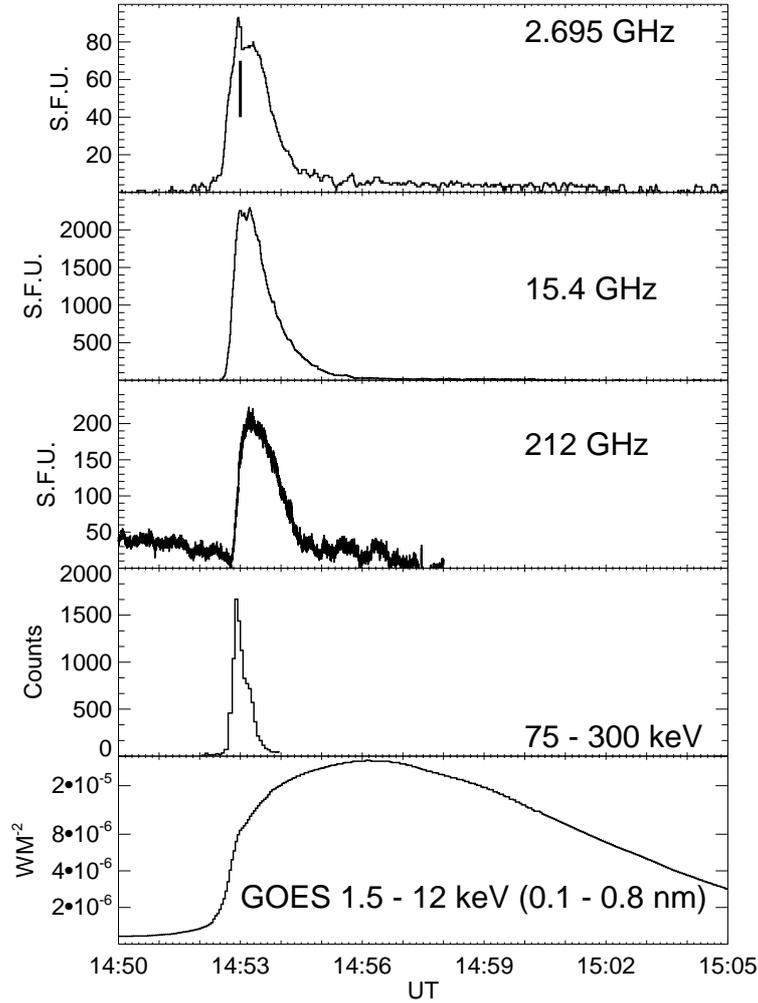}}}
\caption{Radio data at selected frequencies. The vertical bar on the 2.695~GHz
plot indicates the short pulse. We include also 75--300~keV
HXR from RHESSI and 1-8~\AA\ SXR from GOES.}
\label{fig:profiles}
\end{figure}

The time profiles are shown in Fig \ref{fig:profiles}. 
The flare appears as a single impulsive peak in GOES 
data with a start at around 14:50~UT, peak time at 14:56~UT 
and a total duration of about 20 minutes. The soft-X 
rays (SXR) emission has a maximum flux $2.9\times 10^{-5}\ 
\mathrm{Wm}^{-2}$ (M2.9).  We have derived a temperature 
$T_{goes}=14.4$~MK and emission measure 
$EM_{goes}=3.4\times 10^{48} \ \mathrm{cm}^{-3}$ 
using the Chianti 6.0.1 Coronal Model, during the HXR impulsive 
peak interval (14:52:50 - 14:53:00). \\ 

The impulsive phase is clearly visible at all radio frequencies covering 
almost two decades from 1 to 200 GHz, starting at approximately 14:52:30~UT.  
A short pulse is observed at the beginning of the event between 14:52:50 and
14:53~UT (indicated by a vertical bar on the 2.695~GHz panel) with a flux in excess of the main
emission well defined between 1.415 and 4.995~GHz, above this range it is hardly
distinguishable. The short duration and the narrow spectrum of this pulse 
reminds us of a similar one that also occurred during the rising phase 
of an event ({\tt SOL2002-08-30T13:28}, \opencite{GimenezdeCastroetal:2006}).  
At 212 GHz, the flux density time profile is composed of a single peak, with 
maximum at 14:53:20~UT with a total duration of around 2~minutes.
There is no clear evidence of an extended phase like in 
\inlinecite{Luthietal:2004a} and \inlinecite{Trottetetal:2011}.  Besides,
the HXR emission above 50 keV ends around the peak time of the radio emission, 
although the start time of both are similar, below 30 keV the emission
extends longer.  We note that this is one of the weakest GOES events that 
has a clear submillimeter counterpart. Indeed, as a comparison, the flare 
{\tt SOL2002-08-30T13:28} (X1.5) had a 212 GHz peak density flux of 
150 s.f.u.  \cite{GimenezdeCastroetal:2009}  \\

\begin{figure}
\centerline{\resizebox{0.8\textwidth}{!}{\includegraphics{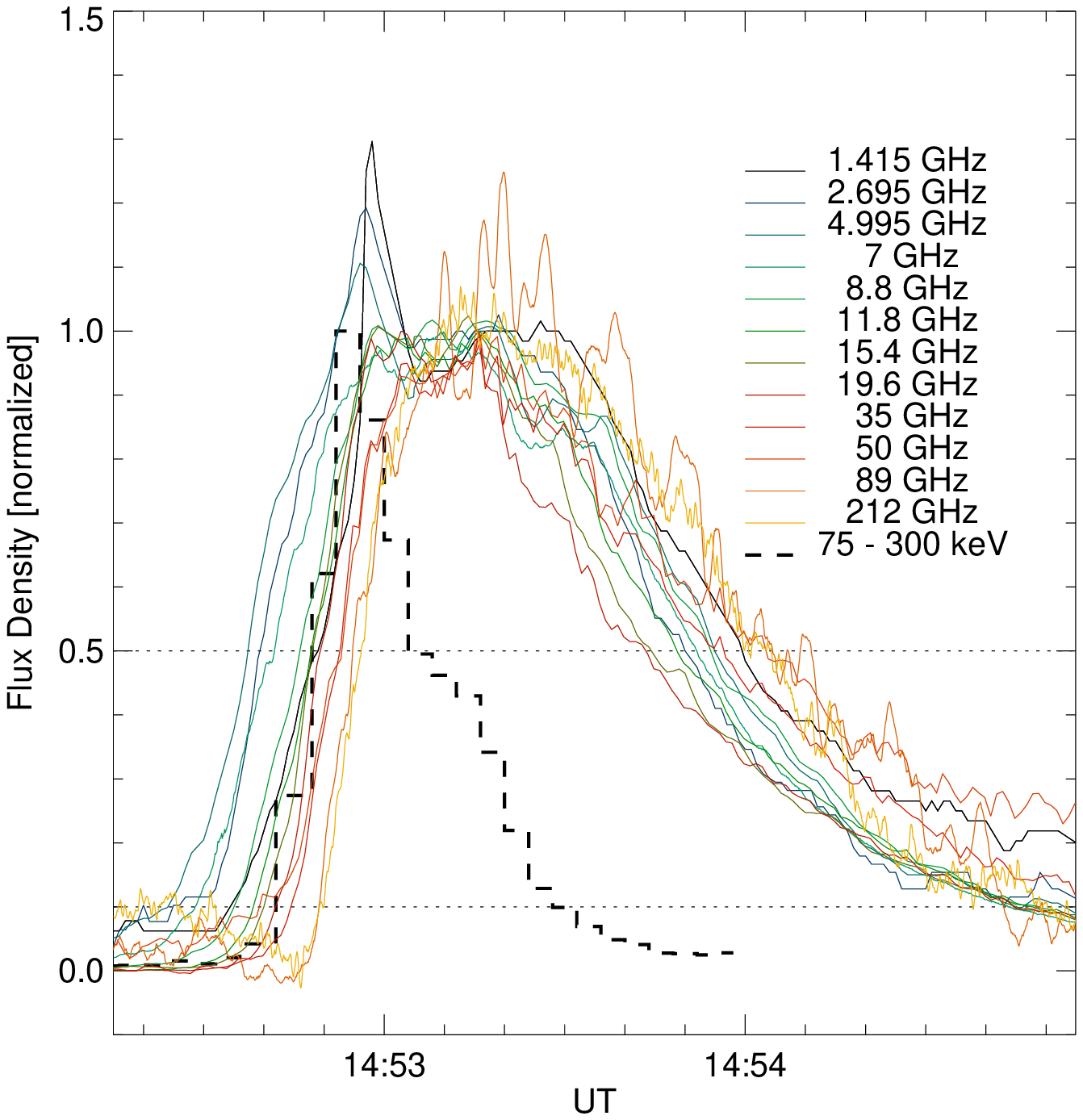}}}
\centerline{\resizebox{0.8\textwidth}{!}{\includegraphics{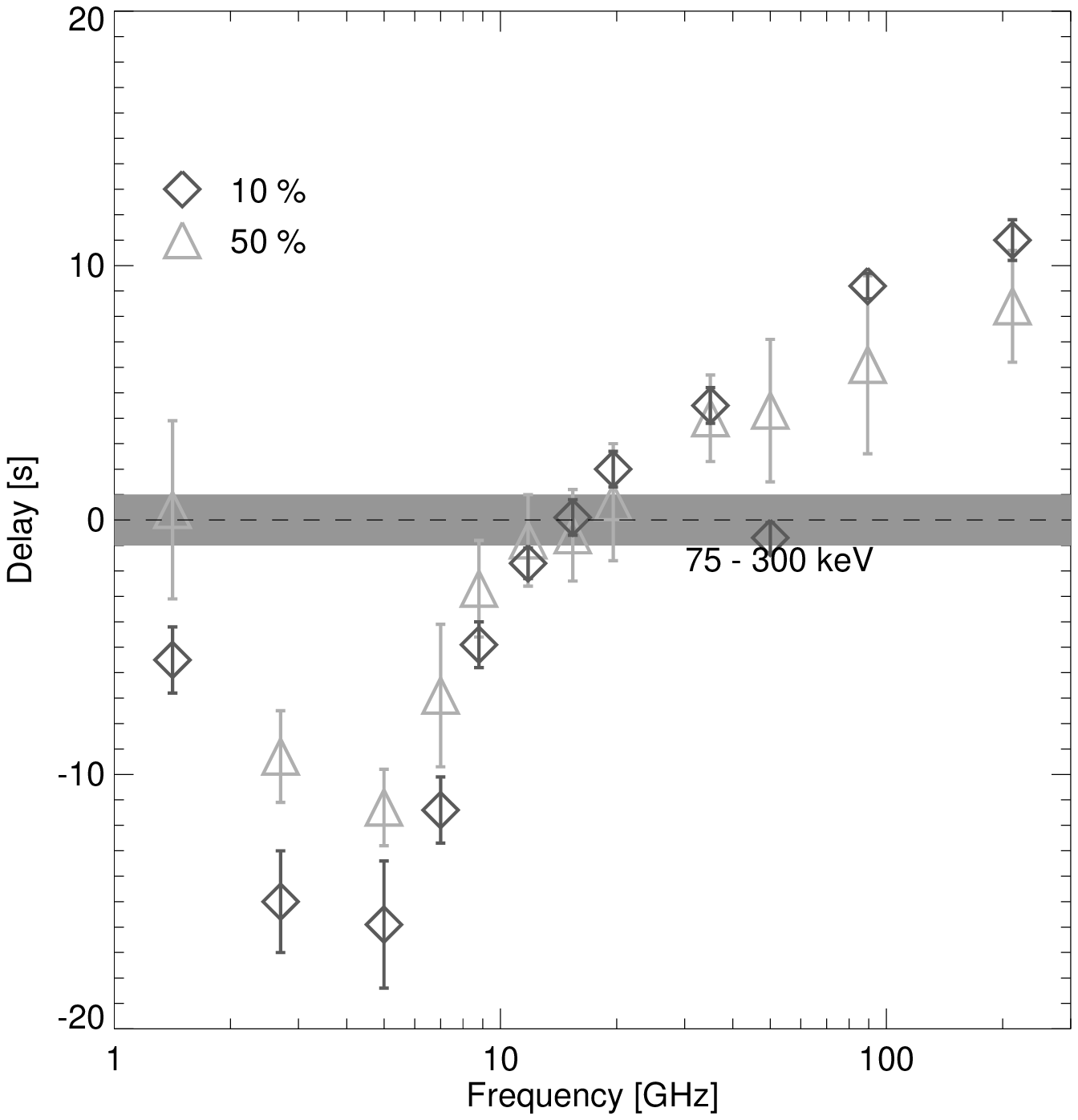}}}
\caption{Top: normalized flux densities from 1.415 to 212 GHz. Normalization
fluxes are taken in the interval 14:5310 -- 14:5330 UT. The dashed
curve is the normalized 75--300~keV flux. Dotted horizontal lines, represent 
10\% and 50\% levels.  Bottom: Measured delays between radio and HXR.  The horizontal 
gray band represents the uncertainty at 75--300~keV rising levels.}
\label{fig:delays}
\end{figure}

We observe a frequency dependent time delay in the radio data, 
which in the four second integrated HXR data is not evident. 
Normalized time profiles integrated in one second bins 
are shown in Figure \ref{fig:delays} (top) at 
each frequency with different colours.  
This figure gives the impression that each different frequency 
starts at slightly different times. To further investigate this, we 
analyzed the rising times at different frequencies. Since 
the start time is subjected to big uncertainties, we compared the 
time elapsed  between a reference time and the time at which 
the emission reaches a certain normalized flux density. 
Moreover, we find more accurate to measure delays during the rising 
phase than  during the, almost flat, maximum.  We used two
different normalized levels, 10\% and 50\%.  For each 
frequency $\nu$ we found $t_{10}(\nu)$ and $t_{50}(\nu)$, 
the times when the flux density reaches the 10\% and 50\% 
relative level respectively. The reference is the 75--300~keV 
emission, with $t_{10}^*$ and $t_{50}^*$ defined when
the HXR flux is 10\% and 50\% of the peak flux respectively. 
Delays are then defined as $\Delta_{10}(\nu) = t_{10}(\nu)-t_{10}^*$ and 
$\Delta_{50}(\nu) = t_{50}(\nu)-t_{50}^*$. To some
extent $t_{10}^*$ and $t_{50}^*$ are arbitrarily defined, 
but in doing so we simultaneously have an indication of the 
relation between radio and HXR.  Delays are determined with 
0.5~s accuracy, derived from the worst time resolution we have 
in the radio data.  The result is shown in Figure \ref{fig:delays} 
(bottom). Above 10~GHz there is a continuous shift, which is, 
within the uncertainties, independent of the level (10\% or 50\%).
This is qualitatively observed in the time profiles (Fig. 
\ref{fig:delays}, top).  Previous works have shown that 
millimeter/submillimeter emission is delayed from microwaves
(see e.g. \opencite{Limetal:1992},\opencite{Trottetetal:2002},
\opencite{Luthietal:2004a}), but this is the first time to 
our knowledge that a {\em spectrum} of the delay is presented. 
Below 7~GHz the presence of the short pulse distorts this trend. \\

\subsection{X-Ray imaging and radio-source positions}
\label{sec:images}

\begin{figure}
\centerline{\resizebox{8.95cm}{!}{\includegraphics{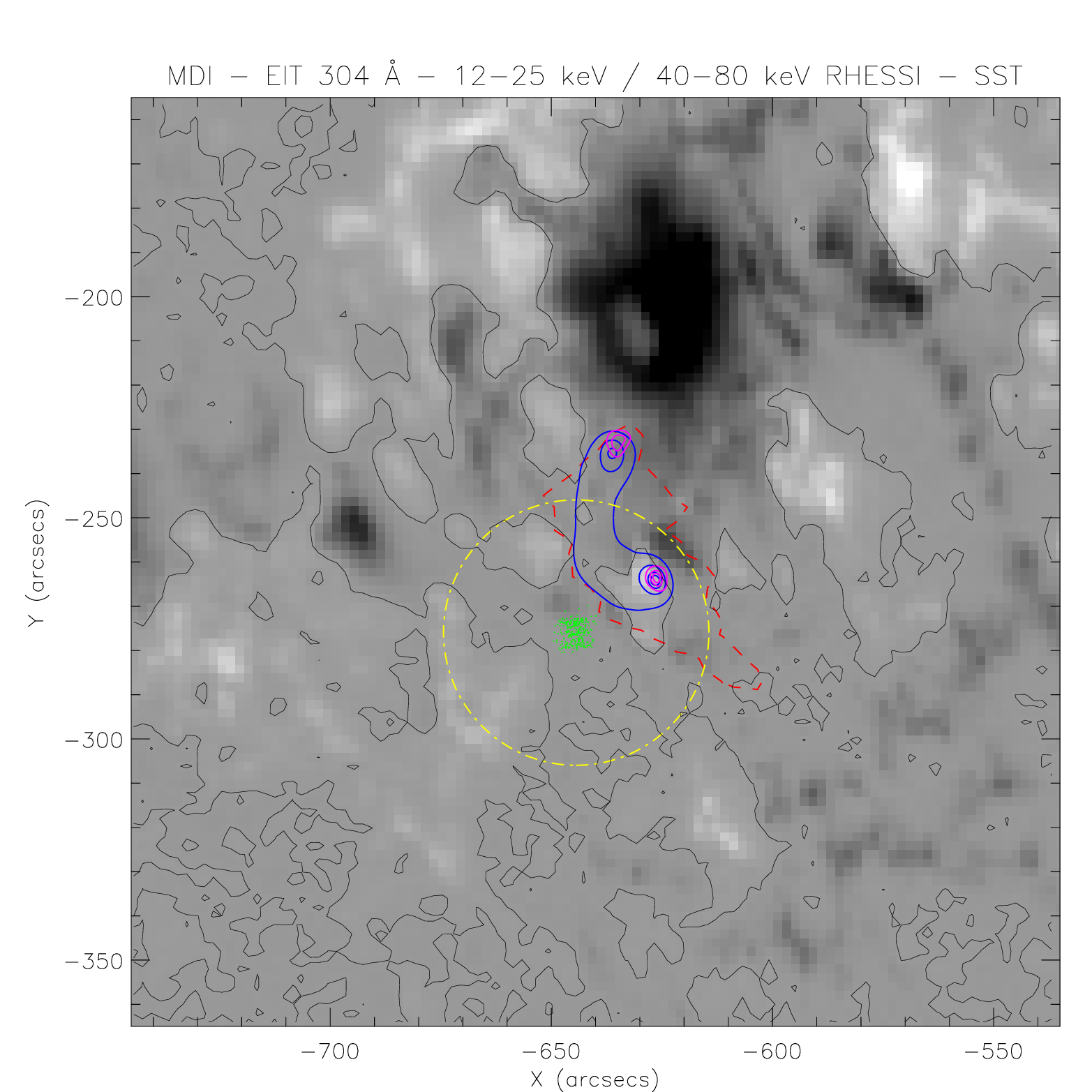}}}
\caption{MDI magnetogram closest in time to the event. The black lines denote
magnetic polarity inversion lines. Blue contours are RHESSI 12--25~keV 
for the interval 14:52:52 to 14:52:56 UT, magenta 
lines are RHESSI 40--80~keV contour levels at 20\%,
50\% and 80\% of the image maximum for the same time interval. 
A contour corresponding to 60\% of the maximum intensity of the closest in 
time EIT image in 304~\AA~is also shown (red dashed contour). Green points 
indicate the centroid positions of the submillimeter emission
from 14:52:40 to 14:54:40 UT (impulsive peak in radio) averaged every 0.4~s. 
The yellow dot--dashed circle denotes the absolute uncertainty in the 
determination of the radio--source positions ($\sim$30\arcsec), mainly 
limited by pointing accuracy.}
\label{fig:MDI-HXR-sst}
\end{figure}

We produced X-ray images using RHESSI data. These images are
constructed with the {\tt PIXON} algorithm
\cite{Hurfordetal:2002} with an accumulation time of four seconds (from
14:52:52~UT to 14:52:56~UT) for lower energy bands (below 80~keV), and
twelve seconds (from 14:52:52~UT to 14:53:04~UT) for higher energy bands (above
100~keV). We used collimators 1--6 and a pixel size of 0.5\arcsec. We considered
the following energy bands: 12--25~keV, 40--80~keV, 100--250~keV,
150--300~keV and 250--300~keV. The images show two sources clearly
defined (see Fig.~\ref{fig:MDI-HXR-sst}). There is a loop or arcade 
connecting the two footpoints visible in the 12-25 keV 
energy band image (see Fig.~\ref{fig:MDI-HXR-sst}).\\

Fig.~\ref{fig:MDI-HXR-sst} shows the  magnetogram closest in time
($\sim$14:27 UT) to the flare overlaid by RHESSI contours in the
12--25 (blue contours) and 40--80 (magenta contours) keV energy bands. 
Also included is a 60\% of the maximum intensity contour of the
closest in time 304~\AA\ EIT image (red lines).  Black thin contours correspond to the
magnetic polarity inversion line, {\em i.e.} they separate positive from
negative line of sight magnetic field. It is evident that both
RHESSI footpoints overlay opposite sign polarities, with one of them
located on the positive polarity, the evolution of which we discussed in
Section~\ref{sec:photosphere}, and the other one on the negative
polarity, north of it. Whithin the uncertainties of the reconstruction
method, we did not observe any displacement in the HXR sources.\\

We determined the centroid of the sources  emitting at 212 GHz
every 40~ms during the impulsive phase of the event,  
assuming that the source size is small compared to 
the SST beam sizes (For a review of the multi--beam method see \opencite{GimenezdeCastroetal:1999}). 
At the same time, we corrected the flux density for mispointing. We note that since 
beam sizes are of the order of arcminutes, when they are not 
aligned with the emitting source the flux obtained from a single 
beam may be wrong. \\

 In Figure~\ref{fig:MDI-HXR-sst} we have superimposed 
the positions of the 212~GHz burst emission centroids averaged every 0.4~s.  
They seem to be separated by 25\arcsec\ from one of the
HXR footpoints.  The dot--dashed yellow circle represents the absolute uncertainty in the 
determination of the radio source position ($\sim$30''); this uncertainty is 
mainly due to the radiotelescope pointing accuracy. Because of this
large absolute uncertainty, it is not possible to determine 
where the submillimeter source is located inside the loop, at the 
footpoints or at the loop top.  The two X-ray sources in the flare seem to 
be associated with the bright ultraviolet enhancement seen by EIT encircling them. This
brightening would correspond to an upper chromospheric loop also
traced by the lowest energy RHESSI isocontours. Timing analysis does
not reveal a displacement of the 212~GHz emission centroids in a
privileged direction during the impulsive phase of the event. 
We can infer the compactness of the submillimeter
source or sources because of  the low level of spread,
lower than 10\arcsec. \\

\section{Spectral Analysis}
\label{sec:analysis}

\begin{figure}
\centerline{\includegraphics{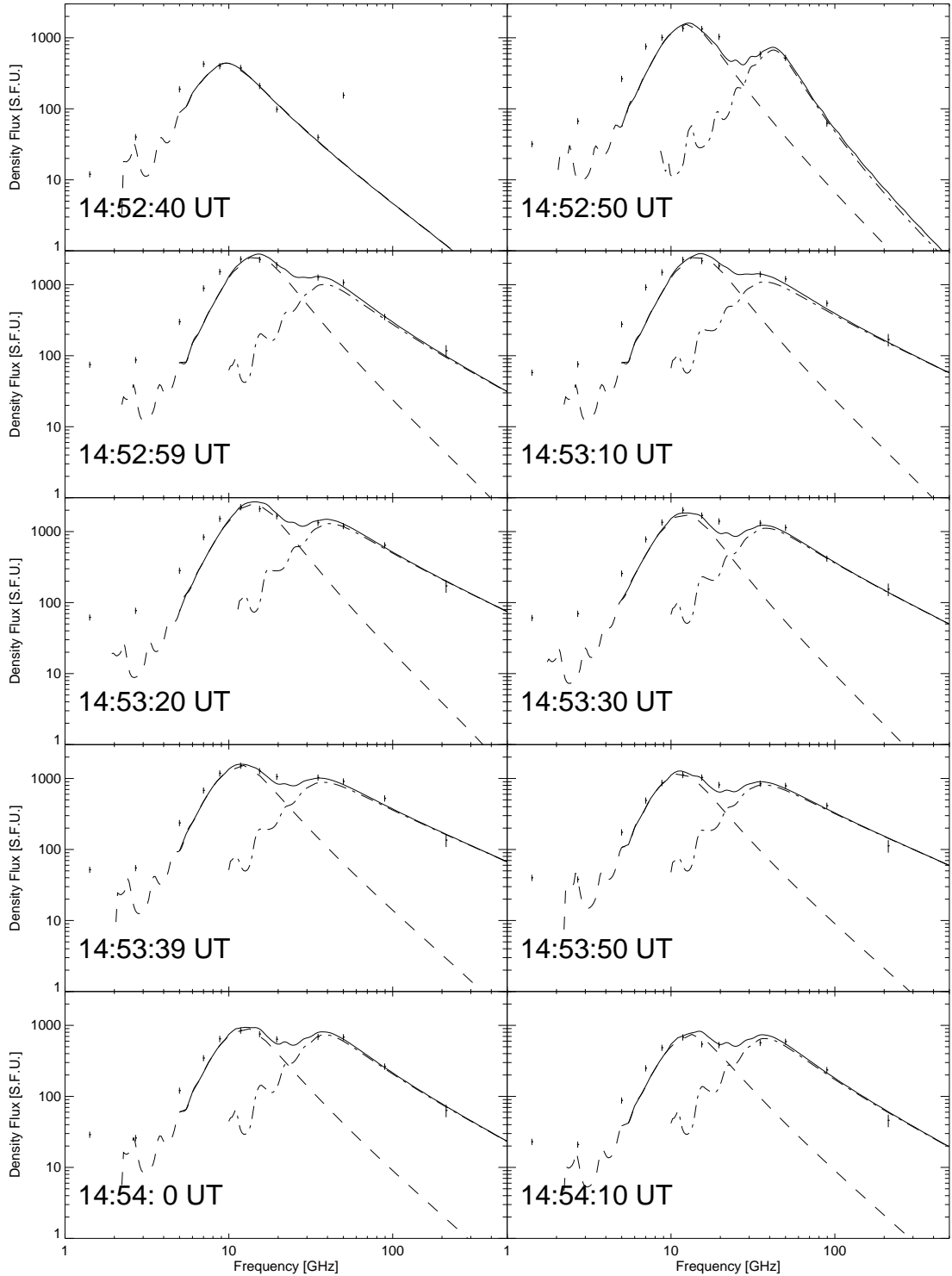}}
\caption{Radio spectra ad selected times.  The data, integrated in 1~s bins, are
represented with their error bars. Dashed curves correspond to the low frequency fitting and
dot-dashed curves to the high frequencies fitting. Continuous lines represent
the total solution (low frequency + high frequency). }
\label{fig:rspec-fit}
\end{figure}

\begin{figure}
\centerline{\resizebox{!}{0.5\textheight}{\includegraphics{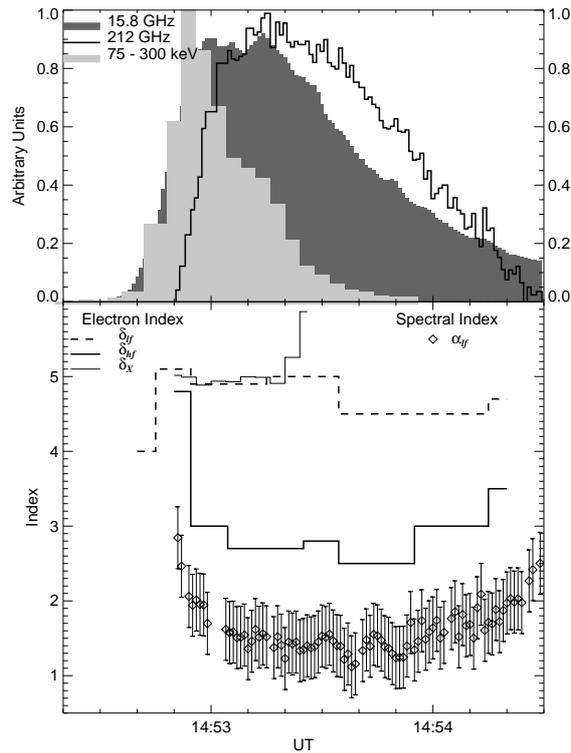}}}
\caption{Top panel: time profiles at 15.8 GHz (dark gray), 75--300~keV
(light gray) and 212~GHz (continuous line). The bottom panel shows the 
time evolution of the electron indices together with the values and error 
bars of the high frequency optically  
thin spectral index $\alfahf$.  }
\label{fig:delta}
\end{figure}

Figure \ref{fig:rspec-fit} shows integrated radio spectra at selected one 
second time intervals. The existence of composed spectra is evident 
after 14:52:50~UT.  We distinguish two components that we call the 
{\em low frequency} and {\em high frequency} components. The low frequency 
one has a peak frequency around 10~GHz, while the high frequency is maximum 
at around 35~GHz. Because only the high frequency component shows clearly 
its optically thin part, we have used frequencies 50, 89.4 and 212~GHz to 
compute the spectral index $\alpha_{hf}$ along the event. Figure \ref{fig:delta} 
bottom shows $\alpha_{hf}$ in function of time together with error bars. 
During the maximum of the event the spectral index remains stable, but
at the beginning and the end a softening is observed, {\em i.e.}  the index
has an SHS behaviour. To get insight into the characteristics of electron 
populations that produced this emission, we have fitted the data to two 
homogeneous gyrosynchrotron sources using the traditional \inlinecite{Ramaty:1969} 
procedure. The suprathermal electron distributions are represented by a power law with
electron indices $\deltalf$ and $\deltahf$ for the low frequency and 
high frequency sources respectively. At each time interval, a model was fitted 
to the low frequency and another model to the high frequency, the sum of both 
was compared to the data until the best solution was obtained. In general we 
fixed the source size, the magnetic intensity, medium density and the electron 
energy cutoffs (see Table \ref{tbl:parametros}). We allowed changes only in 
the electron indices $\deltalf$, $\deltahf$ and the total number of 
accelerated electrons.  The fittings are shown in the Figure \ref{fig:rspec-fit}.
It is evident that below 9~GHz the fittings are rather poor, which is 
an indication that the source is not totally homogeneous \cite{Kleinetal:1986},
nonetheless we may be confident on the electron index and number, which
depend on the optically thin part, and on the magnetic field which is defined 
by the peak frequency. Moreover, the good agreement with similar parameters
derived from HXR is another indication of the goodness of the fittings 
(see below).\\

\begin{table}
\begin{tabular}{l|l|l|l}
                & \bf Low Frequency & \bf High Frequency & \\
\bf Parameter   & \bf Component     & \bf Component      & \bf Unit \\
\hline
Mag. Field      &   380             &     2000           & G \\
Diameter        &   18              &     5              & arc sec\\
Height          &   $10^9$          &    $10^8$          & cm \\
Low En. Cutoff  &   20              &     20             & keV\\
High En. Cutoff &   10              &     10             & MeV\\
Maximum Total Number of $e^-$& $4.7 \times 10^{37}$ &  $7.6 \times 10^{34}$ & \\
\end{tabular}
\caption{Fixed parameters used to fit the spectral data and maximum total electron number derived
from the computations.}
\label{tbl:parametros}
\end{table}

\begin{figure}
\centerline{\includegraphics{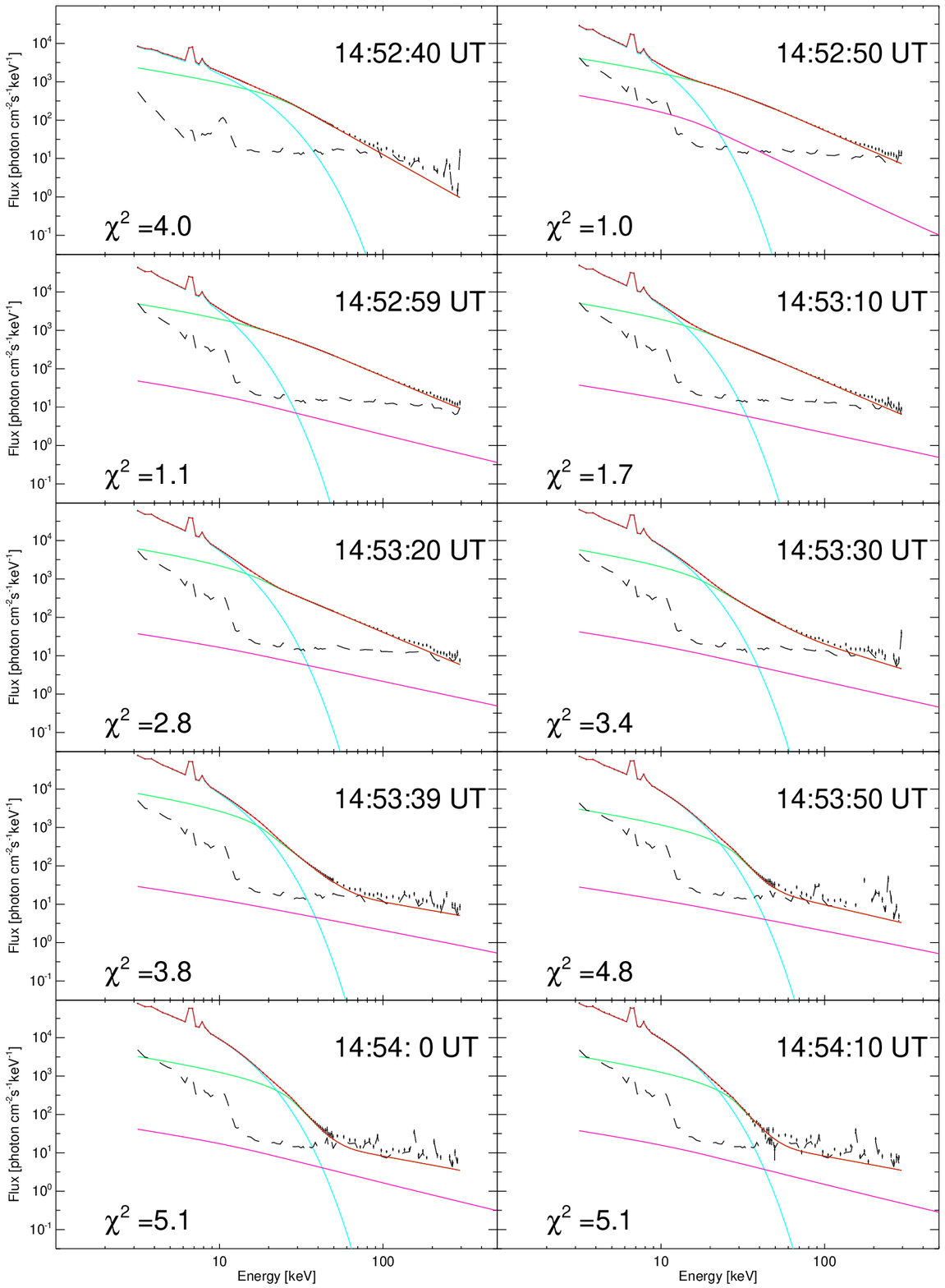}}
\caption{ HXR spectra at selected interval times. Data are integrated in 4~s 
bins and represented with error bars. Blue curves represent the thermal solution,
green curves are the broken power law solution and red curves are the total solution
(Thernal+broken power law).  Magenta curves represent the expected emission from the
relativistic electrons used to fit the radio high frequency component.
Dashed curves are the background.}
\label{fig:xspec-fit}
\end{figure}

HXR spectra were taken from 14:52:00UT to 14:54:00UT in 4 second
intervals, using front detectors 1, 3, 4, 5, 6, 8 and 9, for energies 
between 3 and 290~keV. We excluded the time bin when there was a change 
in attenuator state, from 0 to 1. Due to the high flare activity during 
this RHESSI observation interval, we selected the background emission 
from the subsequent {\em night} period (15:18:48 -- 15:22:12~UT).  Figure 
\ref{fig:xspec-fit} shows the photon flux spectra at selected time intervals 
along with the fitting used to calibrate the data. Using standard 
{\tt OSPEX}\footnote{see 'OSPEX, Reference Guide', Kim Tolbert, at
\url{http://hesperia.gsfc.nasa.gov/ssw/packages/spex/doc/ospex_explanation.htm}} 
procedures we applied a model of a thermal source, including continuum and 
lines, and a double power law.  We found significant counts up to 300~keV, 
during peak time (14:52:59~UT). The thermal component is well represented
by an isothermal source with a mean temperature $T_{hsi}=20.7$~MK and
emission meassure $EM_{hsi}=1.2 \times 10^{48} \ \mbox{cm}^{-3}$ during the HXR impulsive
peak interval (14:52:50 - 14:53:00).  The break energy remains always below 
80~keV, and the electron index below it is always harder than the index
above. The later was used to compare with radio data because electrons with 
energies $< 100$~keV affect the gyroemission only in the optically thick 
part of the spectrum \cite{Whiteetal:2011} which we did not try to fit as noted
before.  We also have to take into account that HXR emission depends on the 
electron flux, therefore for non relativistic particles we should add 0.5 to 
$\delta_X$ to compare with radio \cite{Holmanetal:2011}, and, since the HXR 
spectra are computed below 300~keV, we should compare $\delta_X$ 
with $\deltalf$, because relativistic electrons are needed to produce gyrosynchrotron 
emission for frequencies above 50~GHz.  In the bottom panel of Figure \ref{fig:delta} 
we can see the evolution of electron indices. We observe that $\delta_X$ 
(dot-dashed curve) remains stable at the beginning and increases at the 
end. The low frequency index $\deltalf$ remains stable until 14:53:25~UT and 
is comparable to $\delta_X$, then a sudden change takes place making it 
harder. On the other hand, the high frequency index (continuous line) $\deltahf$
is much harder than $\deltalf$ and $\delta_X$ as it was observed 
in previous works (see e.g. \opencite{GimenezdeCastroetal:2009}, \opencite{Trottetetal:1998}) 
and it evolves in the same way as the spectral index $\alpha_{hf}$.\\

\section{Discussion}
\label{sec:discusion}

The optically thin emission above 50~GHz is produced by relativistic
particles, while the optically thick emission at microwaves 
and the HXR observed by RHESSI are emitted by mildly relativistic 
particles.  We roughly divide the analysis in these two energy bands.

\subsection{Dynamics of mildly relativistic electrons}
\label{sec:mild-relativistic}

In order to understand the dynamics of the mildly relativistic
electrons we compare the HXR emission, which it was observed
up to 300~keV with the {\em low frequency} radio data. 
The comparison of the temporal evolution of both sets of data
(see Figures \ref{fig:profiles} and \ref{fig:delta}) 
supports the existence of trapped electrons because: 1) 
the duration of the impulsive phase in HXR is shorter
than in radio and 2) the peak time in HXR occurs before the
radio peak, even at low frequencies. 
Therefore, the HXR time profile is not necessarily the 
representation of the injected electrons, since 
there are transport effects along the loop,
or at least, HXR may represent the injected electrons
that precipitate directly, without being subject to trapping.\\

We can use the spectral analysis to derive the rate 
of the injected electrons in the emitting area in function
of time. To do so, we write a simplified continuity 
equation that depends only on time, since we are not
interested on how the electron distribution changes in
terms of energy, pitch angle, or depth. In this simplified
model, we are interested only in the total instantaneous 
number of electrons, $N(t)$, inside the magnetic loop, 
incremented by a source, $Q(t)$, from the acceleration
site and decremented by the precipitated electrons, $P(t)$.
Therefore, the continuity equation should be 
\begin{equation}
\frac{dN(t)}{dt} = Q(t) - P(t) \ . \label{eq:cont}
\end{equation}
Integrating the above equation in the interval $(t_i,t_{i+1})$
(with $i=0,1,2\dots$) and solving for $Q(t_i)$ yields
\begin{equation}
Q(t_i)\Delta t_{i+1} = N(t_{i+1}) - N(t_i) + P(t_i)\Delta t_{i+1} \ ,
\label{eq:injection}
\end{equation}
with $\Delta t_{i+1} = t_{i+1} - t_i$.  Since we are comparing 
$< 300$~keV emission with gyrosynchrotron, we can identify the instantaneous 
number of electrons inside the loop $N(t)$ with the trapped particles 
emitting the lower frequency component.  On the other hand, the particles leaving the 
volume $P(t)$ produce the HXR emission observed by RHESSI. In our picture, the 
low frequency component is produced all along a loop with a length of 
$10^9$~cm, while the HXR emission is produced in a narrow slab (see {\it e.g.} 
\opencite{Holmanetal:2011}) with a very small surface (see Figure~\ref{fig:MDI-HXR-sst}); 
therefore, we can neglect the gyroemission produced within this small volume.
Furthermore, no change would be appreciated if one includes the particles
responsible for the high frequency component since they are two orders of magnitude
less than those that produce the low frequency component. (See Table 
\ref{tbl:parametros}) \\

\begin{figure}[h!]
\centerline{\resizebox{!}{0.5\textheight}{\includegraphics{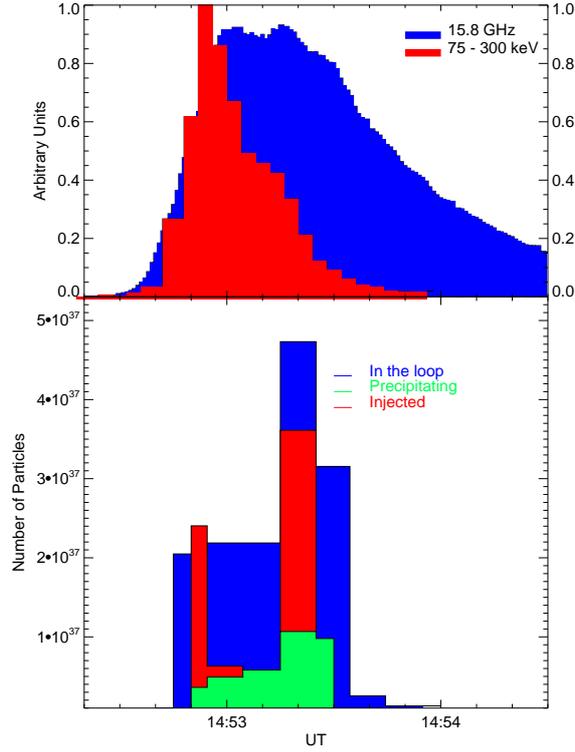}}}
\caption{Top: time profiles at 15.4 (blue) and 75--300~keV (red).
Bottom: the precipitated electrons $P(t)\Delta t$ (green);
the instantaneous number of electrons, $N(t)$ (blue), and
the injected electrons $Q(t)\Delta t$ (red).}
\label{fig:injection}
\end{figure}

We divided the event in 10 second intervals
and assumed that within these intervals the conditions do not change. 
The computation of Equation~\ref{eq:injection} is straightforward and 
the result is presented in Figure \ref{fig:injection}.  Since 
$> 100$~keV data have good S/N ratio only between 14:52:50~UT and 14:53:30~UT, 
(see Figure \ref{fig:xspec-fit})
the analysis is restricted to this interval, although is clear from the
time profiles that there are emitting particles before and after.
We observe a continuous injection with two peaks, one at the 
beginning and the second during the decay of the impulsive 
phase. To verify our results, we sum the precipitated electrons, 
$\sum_i P(t_i)\Delta t=3.4\times 10^{37}$,
and we compare this number with the maximum number of  
electrons existing instantaneously inside the 
loop, $\max[N(t_i)] = 4.7\times 10^{37}$. The 
difference between these two numbers maybe due to the fact 
that we are limited to the time interval in which the HXR data are 
statistically meaningful; therefore, we cannot track the precipitation 
until the end of the gyrosynchrotron emission.\\
 
We observe that changes in $\deltalf$ (Fig. \ref{fig:delta}) 
can be related to the injections occurred at around 14:52:51~UT and 
14:53:15~UT (Figure \ref{fig:injection}).  The 
electron index $\delta_X$ lies around 5, as $\deltalf$ for the same 
period, although, it increases slowly until 14:53:20~UT  when it 
suddenly softens by around 0.5. The difference in time evolution of 
$\delta_X$ and $\deltalf$ is an indication that the softening of the 
former is a consequence of the trapping.  Since $\deltalf$ has two 
constant values, we conclude that it is not affected by the medium.
In the above analysis we rely upon the parameters derived from
the radio and HXR data fittings.  Although we do not claim that 
the obtained solutions are unique, the fact that two independent
fittings give very comparable results, give us confidence on them.  \\

The progressive delay of the radio emission observed in 
Figure \ref{fig:delays} must be interpreted differently depending 
on the frequency range.  Between 1 and 5 GHz, the short pulse dominates 
the emission during the rising phase; therefore, its contribution should 
be removed to asses the delay of the main emission. This may lead to 
ambiguous results, hence, we preferred not to analyze delays in this range.  
In the range between 9 and 30~GHz lies the peak frequency, i.e. the optical 
opacity $\tau$ is approximately 1, hence, the gyrosynchrotron self absorption 
is critical.  The peak frequency ($\tau\simeq 1$) of the low frequency 
component shifts from approximately 7~GHz to 15~GHz from 14:52:40~UT to the 
peak time around 14:5259~UT (Fig. \ref{fig:rspec-fit}).  Since there is no 
change in the magnetic field, this can be interpreted by the accumulation 
of accelerated electrons inside the loop due to the trapping that increases 
its density, which is the dominant factor of the self-absorption mechanism. 
The shift makes the low frequencies more absorbed and thus increases the 
relative importance of higher frequencies.  Therefore, even with a rather 
constant $\deltalf$ (Fig. \ref{fig:delta}) the progressive delay should 
be observed. In our fittings at 14:52:40~UT we have $\tau(\nu=7\ \mbox{GHz})=0.7$ and 
$\tau(\nu=10\ \mbox{GHz})=0.15$. Later on, during peak time
$\tau(\nu=7\ \mbox{GHz})=60$ and $\tau(\nu=10\ \mbox{GHz})=3.7$. 
We note that the emission is proportional to $1-e^{-\tau}$, hence, 
we have a relative amplification of 
$(1-e^{-60})/(1-e^{-0.7}) \simeq 1.4$ at 7~GHz, while it 
is $(1-e^{-3.7})/(1-e^{-0.15})\simeq 7.5$ at 10~GHz. The relative
amplification changes the rate at which the emission rises and, therefore,
the time when the signal reaches a certain level with respect to its maximum.\\

\subsection{Relativistic Electrons}
\label{sec:relativistic}

The centroid position of the emitting source at 212~GHz remains quite
stable during the flare which may imply that the source is compact. 
Moreover, it is placed 25\arcsec~far from one of the HXR footpoints 
(Figure \ref{fig:MDI-HXR-sst}).
A similar result was obtained by \inlinecite{Trottetetal:2008} for the
{\tt SOL2003-10-28T11:10} flare; during the impulsive phase (interval B
in their work) the centroid positions of the 210~GHz emission lie at 
approximately 10\arcsec\ from the center of one of the  HXR 
footpoints (250--300~keV), but are coincident with the location of
precipitating high energy protons with energies above 30~MeV seen
in $\gamma$-ray imaging of the 2.2~MeV line emission. Since for our work 
we do not have $\gamma$-ray imaging to compare with, we should be 
cautious because the uncertainty in position is of the order of the 
position shift. \\

Since we observe the optically thin part of the high frequency 
spectra, the obtained $\deltahf$ is not affected by the medium and,
within data uncertainties, it must be correct.  On the other
hand, we assumed a standard viewing angle $\theta=45^\circ$ which gives 
us a mean value of the magnetic field $B$ and total number of electrons
$N$.  Increasing (reducing) $\theta$ results in smaller (larger) $B$ and 
$N$. Although we cannot rightly evaluate $\theta$ with our data, 
we do not expect an extreme value for it since the AR is located
not far from Sun center (E43). Furthermore we did consider an isotropic
electron distribution. The total number of accelerated electrons 
is $8\times 10^{34}$ during peak time, and they should not produce enough
bremsstrahlung flux to be detected by RHESSI detectors.  We confirmed this by computing
its HXR emission using the {\tt bremthick}\footnote{Developed by G. Holman, last revision May 2002. 
Obtained from the RHESSI site: \url{http://hesperia.usfc.nasa.gov/hessi/modelware.htm}}
program (magenta curves in Figure \ref{fig:xspec-fit}) which is orders 
of magnitude smaller than the mildly relativistic electron emission and 
remains below (except during one time interval) the background. \\

The spectral index of the high frequency component $\alpha_{hf}$ shows 
a SHS behaviour, but in this case  we cannot conclude whether its origin
comes from the acceleration mechanism or from the interaction with the medium 
as before. We tend to think that the former should be the cause, since these
are relativistic particles and their interaction with the medium should 
be less effective.  The emission must come from a compact region with a 
strong magnetic field and the electron index $\deltahf$ should be 
harder than $\delta_X$ and $\deltalf$, as is the case. These arguments 
support the evidence of the existence of a separated source where 
relativistic electrons are the responsible for the emission.\\

The progressive delay of the radio emission above 50~GHz 
(Figure \ref{fig:delays}) can be interpreted considering the initial 
hardening of the spectral index $\alpha_{hf}$. If it is due to the 
acceleration mechanism that accelerates first the {\em low} energy particles 
and later the {\em high} energy particles, then, the progressive 
delay is a consequence. 

\section{Final Remarks}
\label{sec:fin}

From our simple continuity model, we inferred the evolution of
injected number of electrons, appearing as a continuous injection with
two distinctive pulses separated by approximately 30 seconds. From the
timing of those pulses, the first one produces the
HXR and initiates the radio low frequency emission. The
second pulse, while slightly stronger than the first, builds up into
the radio emission but do not contributes to generate more HXR.
This description suggests that the two injection might have
different initial pitch angle distributions, because our fittings 
do not show magnetic field changes during the burst. The first injection
might be formed by an electron beam aligned with the magnetic field
direction, a fraction of the population is trapped by magnetic mirroring, while
the other fraction enter the loss cone and precipitates, producing
HXR. The second pulse should then be formed by a beam with a wider
pitch angle distribution, or even isotropic, keeping most of
the electrons trapped, producing radio emission, with little 
precipitation (no significant increase in HXR).\\

From our observations we conclude that the radio spectra cannot be
explained by an homogeneous gyrosynchrotron source, therefore we adopted 
a model based on two homogeneous sources since it is the simplest hypothesis that
fits reasonably the data. Although we don't have images to support 
this assumption, it is plausible that it is the case.  As shown by the magnetic 
field evolution discussed in Section 2.2, we can conclude that the flare originates by 
the interaction of magnetic loops anchored in the MMF polarities. Magnetic 
energy is probably increased in the configuration by shearing motions, 
in particular, the rotation of the negative polarity around the positive one.   
After magnetic reconnection occurs, two sets of reconnected loops 
should be present. In the example we have 
analyzed, we observe a set of loops in SXR (and also in EUV) with two HXR 
footpoints. Considering our description in
Section~\ref{sec:photosphere}, we speculate that the second set of 
reconnected loops (not visible) could be anchored in the 
higher magnetic field positive polarity located north of the northern HXR 
footpoint and the negative polarity that rotates around the positive bipole 
polarity. Such a set of reconnected loops would have a larger 
volume than the ones that are observed; therefore, considering that the 
same amount of energy is injected in both sets, the emission in the 
second one could be less intense and, then, not visible.  
This second magnetic structure is also suggested by 
the 25\arcsec\ displacement of the emitting source at 212~GHz respect 
to one of the HXR footpoints (Figure \ref{fig:MDI-HXR-sst}).
While it has been shown that the reconnection of many different magnetic 
loops is more efficient to accelerate high energy electrons (see 
\opencite{Trottetetal:2008} and references therein), we  might conclude that 
this complex mechanism operates even in medium size events as the one 
we analyzed in this work. \\
 
Another possible scenario is a loop 
structure where the {\it low frequency} source represents the coronal
part of the loop (with a lower effective magnetic field strength), and
the {\it high frequency} source represents the low coronal or
chromospheric footpoints of the loop (with a higher effective magnetic
field strength). \opencite{Melnikovetal:2011}, using simulations of
electron dynamics and gyrosynchrotron emission in a loop structure,
demonstrated that two spectral components can be produced from one
single loop, where mildly-relativistic electrons produce microwave
emission in the loop, while relativistic electrons produce higher
frequency (reaching sub-THz frequencies) emission from the footpoints.\\

\begin{acks}
CGGC is grateful to FAPESP (Proc. 2009/18386-7).
CHM acknowledge financial support from the Argentinean grants UBACyT 
20020100100733, PIP 2009-100766 (CONICET), and PICT 2007-1790 (ANPCyT). 
PJAS is grateful to FAPESP (Proc. 2008/09339-2) and to the European Commission
(project HESPE FP7-2010-SPACE-1-263086).  GDC and CHM  are members of the Carrera del Investigador 
Cient\'{\i}fico (CONICET), CGGC is level 2 fellow of CNPq and Investigador 
Correspondiente (CONICET). 
\end{acks}

\bibliographystyle{spr-mp-sola}

\end{article}
\end{document}